# Two-Dimensional Quasi Periodic Structures for Large-Scale Light Out-coupling with Amplitude, Phase and Polarization Control

**YOAV LIVNEH[1, *], AMI YAACOBI[1, 2], MEIR ORENSTEIN[1]**

[1] *Andrew & Erna Viterbi Department of Electrical and Computer Engineering, Technion-Israel Institute of Technology, 32000 Haifa, Israel*

[2] *Rafael Advanced Defense Systems Ltd., POB 2250, Haifa, 3102102 Israel*

[*] *yoavli@campus.technion.ac.il*

**Abstract**: Chip-scale light-atom interactions are vital for the miniaturization of atomic sensing systems, including clocks, magnetometers, gyroscopes and more. Combining as many photonic elements as possible onto a photonic chip greatly reduces size and power consumption, where the critical elements are those interfacing between the 2D circuit and the 3D vapor cell. We introduce a new design method for large scale two-dimensional converter structures, enabling out-coupling of radiation from the photonic chip into the atomic medium. These structures allow light intensity and phase spatial distribution and polarization control, without external light-manipulating elements. Large, 100x100 μm$^2$ structures were designed generating low divergence optical beams with high degree of circular polarization. Simulations obtain mean circular polarization contrast of better than 30 dB.

## 1. Introduction

In recent years, chip-scale atomic systems, utilizing light-atom interactions, have been successful in reducing dimensions and power consumption in quantum metrology applications such as optical clocks [1,2], magnetometers [3], atomic gyroscopes [4], and measurement of international standard (SI) units [5]. Most of the systems utilizing light-atom interactions for sensing and metrology are based on bulk optical components, such as lenses, polarizers, etc. [2,4,6], thus limiting the miniaturization capability and robustness of the system and increasing the packaging cost. One mitigation for this problem is using integrated photonics, where part or all optical components are implemented on a photonic chip.

Three main approaches for realization of light-atom interaction using integrated photonics have been proposed. While the first two methods, hollow-core waveguides [7,8] and atomic-cladding waveguides [9–11], work well for local, small scale interactions, they lack the scalability required by some atomic systems. Due to their local nature they lack the possibility to influence atoms far from the waveguide, even few tens of microns away. The third proposed method is to employ wide one-dimensional grating couplers to out-couple the light from the two-dimensional chip to the three-dimensional realm of the atomic environment [12–15]. In this case, the optical field is shaped with the proper profile inside the chip, and beamed onto an atomic vapor cell to perform the interaction. These couplers allow for additional types of interactions, for example interaction with, and manipulation of ions [16,17]. When polarization control is required, additional elements, such as off-chip polarizers or on-chip metasurfaces [14], are necessary to facilitate these requirements. In this paper we propose a method for converting micrometer-sized optical modes propagating in planar waveguides to arbitrarily shaped millimeter-sized beams propagating into the atomic vapor, with predetermined polarization (e.g. circular, as required in many atom-light interactions). This is achieved by an optimized design of a two-dimensional quasi periodic structure excited by two orthogonal inputs.

The conceptual implementation of our photonic circuit, designed for out-coupling circularly polarized light from the photonic chip, for example as required by a coherent population trapping (CPT) based miniature atomic clock [2], is shown in Fig. 1a. The principle of operation is as follows: light is input into the chip and then split to two paths, each containing a phase shifter. The radiation in each branch is then expanded to the desired dimensions with a wavefront phase correction element. The two radiation fields then propagate perpendicularly in a highly optimized quasi periodic structure, which out-couples the requested beam. The beam polarization can be controlled using the tunable on chip splitter and phase shifters. For circular polarization the splitter should be set for 50/50 and the phase shifters to $\pm\pi/2$. For linear polarization parallel to one axis, the tunable splitter shall be tuned to transfer all power to one branch. For 45 degrees linear polarization the tunable splitter shall be tuned to 50/50 and the phase shifters to equal phase.

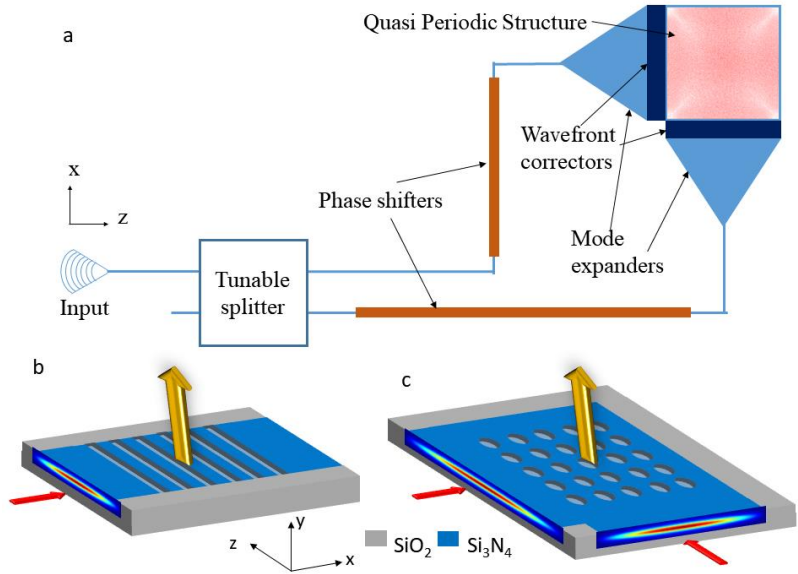

**Fig. 1.** Schematic of (a) suggested full photonic circuit for out-coupling circularly polarized light (not to scale); (b) one and (c) two-dimensional gratings. Electric field mode profiles are superimposed upon the structures at the inputs.

Designing very large structures that can generate useful beams for mm-sized vapor cells present unique challenges. The quasi periodic structures contain many elements, with very strict phase matching requirements. Errors in design or fabrication accumulate over the length of the structure and can lead to poor quality results. This intrinsic sensitivity of the structures leads in turn to tight fabrication tolerance requirements.

The design method proposed here surpasses the traditional method for designing grating couplers [18], by taking into account both the perturbation strength and the loss per unit length of each element. The result is a more accurate design methodology, capable of designing arbitrary field shapes, including polarization profiles. Further constraints can be introduced to the design process, including far field angle and fabrication considerations.

For many light-atom interaction metrology applications, alkali atoms, such as Rb and Cs, are used. The wavelengths required for the interaction are in the range of 700-900 nm. For this work, we focused on the rubidium (Rb) atom $D_1$ transition excited by a wavelength of about

795 nm. Silicon nitride (Si$_3$N$_4$) was selected as the core material of the waveguide system since it has a high optical quality with a moderate refractive index (n = 2) and low losses [19], enabling fabrication of excellent light confining structures. Furthermore, Si$_3$N$_4$ is a standard material in the CMOS industry and is already in industrial use for low loss optical waveguide circuitry. Although the emphasis in this paper is atomic sensors, similar challenges exist in applying 2D photonic circuitry for other three-dimensional sensing, such as biomedical sampling.

The paper is arranged as follows. First the design method is described in detail. This includes description of all preliminary simulations required and other insights. Then simulation results for two types of quasi periodic structures are presented, followed by a discussion of the results and conclusions.

## 2. Design Method

Our goal is to develop a methodology for an optimized quasi periodic structure design for a set of specific requirements. In this work the focus was on a configuration favorable for the atomic sensor, namely relatively large beams emitted perpendicular to the surface with circular polarization. The main challenge is maintaining both the desired output field intensity profile and phase profile while the generating field is propagating in the slab waveguide. To achieve a narrow collimated beam, it is imperative to maintain a constant accumulation of phase per unit cell of the structure, while simultaneously changing the perturbation strength at each unit cell to fit the desired field intensity profile. In the case of perpendicular emission, the exact phase accumulation per cell should be an integer multiple of $2\pi$.

By first examining a one-dimensional grating design method, some insights can be learned and used in the 2D case. Schematic drawings of one- and two-dimensional gratings are shown in Fig. 1b, c, showing two perpendicular propagation directions for the two-dimensional case. A one dimensional grating coupler from a waveguide can be analytically designed to produce an arbitrary field profile. For a desired near field profile $E(x)$ and electric field $A_0$ at the beginning of the waveguide, the requested perturbation strength $\alpha$ at each point along the in-plane propagation direction $x$ is [18]

$$\alpha(x) = \frac{E^2(x)}{2\left[A_0^2 - \int_0^x E^2(t)dt\right]} \quad (1)$$

The most direct and efficient method of creating the grating is by perturbing the waveguide itself, by partly or fully etching the slab. The perturbation strength is controlled by changing the duty cycle [20–23]. However, the two-dimensional case adds another layer of complexity, since some cross coupling effects always occur between the perturbation strengths of the two propagation directions. As a result, any attempt to change the perturbation strength in one direction, affects the perpendicular direction as well. Furthermore, the derivation of Eq. (1) assumed that the only energy loss is via radiation out of the grating, symmetrically up and down. This is not always the case, but can be neglected for short devices, since the accumulated error is small. When designing long, mm scale devices, we cannot assume the power distribution is symmetric. For these reasons it is imperative to develop a more rigorous method of designing 2D structures.

We define here the perturbation strength, $\alpha_{fs}$, as the fraction of power radiated from the slab to free space, per unit length, for each propagation direction, at each point in the quasi periodic structure

$$\begin{cases} I_{fs,x}(x,z) = P_{in,x}(x,z)\alpha_{fs,x}(x,z) \\ I_{fs,z}(x,z) = P_{in,z}(x,z)\alpha_{fs,z}(x,z) \end{cases} \quad (2)$$

where $P_{in}$ and $I_{fs}$ are the propagation direction dependent input power per unit width into the scattering element, and power per unit area radiated to free space from that element, respectively. Energy loss $\alpha_{loss}$ is the fractional loss per unit length and is defined using the expression for the transmission between two adjacent elements:

$$\begin{cases} P_{through,\,x}(x,z) = P_{in,\,x}(x,z) \cdot (1 - \alpha_{loss,\,x}(x,z) p_x) \\ P_{through,\,z}(x,z) = P_{in,\,z}(x,z) \cdot (1 - \alpha_{loss,\,z}(x,z) p_z) \end{cases} \quad (3)$$

where $p$ is the element pitch in each direction and $P_{through}$ is the power per unit width traversing the element, in each propagation direction.

The quasi periodic structure is created by shallow etching elliptical holes onto the top of the $Si_3N_4$ slab and overlaying the structure with $SiO_2$. The ellipse shape was chosen because it provides the two degrees of freedom required for the design, as well as convenience for lithographic definition (e.g. compared to sharp rectangles). A unit cell library has been compiled by simulating periodic arrays of ellipses, each with given ellipse and pitch dimensions, 81 library elements in total. Simulations were performed using commercial 3D Finite-Difference Time-Domain (FDTD) program from Lumerical [24]. Fig. 2 shows side and top schematic views of a typical structure used for generating a library element. The boundary conditions are perfectly matched layers (PML) at the x and y boundaries and periodic in z. The simulated device length is 10 μm, the periodic width is exactly one lateral direction pitch and the $Si_3N_4$ slab thickness is 250 nm, with 2 μm of $SiO_2$ above and below and $Si_3N_4$ layer. Although the etch depth can in principle also be a parameter, varying etch depth will make the fabrication difficult. A 40 nm etch depth was selected to provide a large span of perturbation strengths to allow large mm-scale devices, by enabling weak perturbation strengths. The incidence light source was the fundamental transverse electric (TE) slab mode.

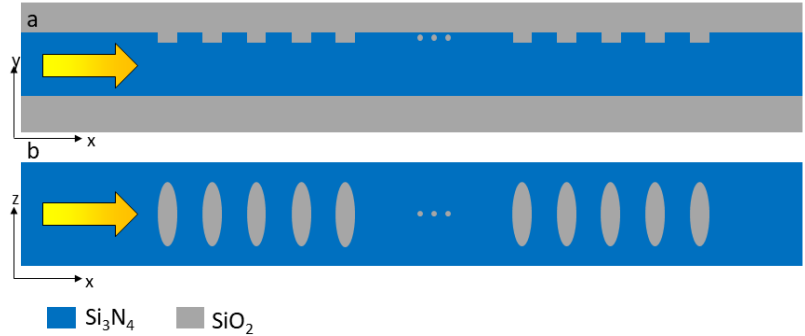

**Fig. 2.** Side (a) and top (b) view of the devices simulated for creating the component library, not to scale. Arrow indicates source injection direction.

The four simulation parameters are the radii $(r_x, r_z)$ in x and z direction, and the pitches $(p_x, p_z)$ in x and z direction. Only the radii are free parameters, since for each radii combination the two pitch values are those required to achieve a far field angle of zero degrees and were found using an optimization algorithm. The algorithm progress was narrated by finding the correct pitch combination for propagation in both directions simultaneously. For each combination of these four parameters, a single library element was simulated using the FDTD method. The input was a z-polarized mode source propagating in the x-direction, and the output field was monitored using a two-dimensional field monitor placed 2 μm above the top $SiO_2$ layer. Trivially, flipping the propagation and polarization directions will yield the same outcome, namely, for the parameter combination $[r_x, r_z, p_x, p_z] = [r_1, r_2, p_1, p_2]$, the $\alpha_{air}$ and $\alpha_{loss}$ values recorded for x-propagating z-polarized input will also be recorded for a z-propagating x-polarized mode, but with swapped indices $[r_2, r_1, p_2, p_1]$.

The perturbation strength per unit length was calculated by taking the transmission at the field monitor in air divided by the actual library element length. Fig. 3a shows an example of the perturbation strength for a z polarized mode propagating in the x direction, using pitch values corresponding to a far field angle of zero degrees. The x-direction pitch values obtained using the optimization algorithm can be seen in Fig. 3b. In both figures, the z-direction values can be found by flipping the images about the $r_x = r_z$ line.

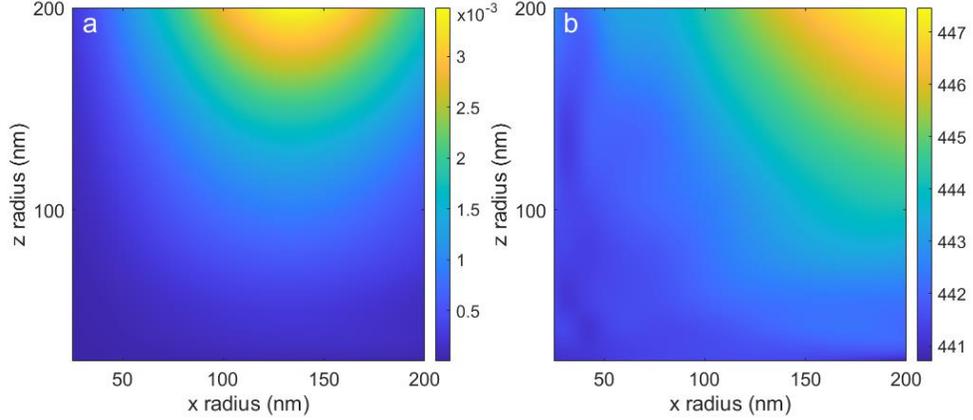

**Fig. 3.** (a) Perturbation strength $\alpha_{fs}$ (µm$^{-1}$) and (b) x-direction pitch values (nm), for light propagating in the x direction, as a function of the ellipse radii

The design methodology is as follows: First, at each (x, z) array coordinate, we calculate the required perturbation strengths for light propagating in either direction, depending on the desired field intensities and the remaining energy in the slab. This is done by solving the following equations for $\alpha_{req}$:

$$\begin{cases} P_{in,x}(x,z)\alpha_{req,x}(x,z) = I_{req,x}(x,z) \\ P_{in,z}(x,z)\alpha_{req,z}(x,z) = I_{req,z}(x,z) \end{cases} \quad (4)$$

$\alpha_{req,i}$ is the required perturbation strength in each direction $i$, and $I_{req}$ is the desired near field intensity at this point. Next, find the radii combination providing the required perturbation strengths, by minimizing the RMS error

$$E = \sqrt{(\alpha_{sim,x} - \alpha_{req,x})^2 + (\alpha_{sim,z} - \alpha_{req,z})^2} \quad (5)$$

where $\alpha_{sim,i}$ is the simulated perturbation strength values. Since only a finite number of radii combinations were simulated in the creation of the library elements, most locations will require two dimensional interpolation of the perturbation strength between known values to find the correct radii combination, and the pitch values will be interpolated accordingly. In the last step, we propagate the field in the slab to the next position using Eq. (3), and perform the sequence again.

This method allows for the design of an arbitrary desired beam shape. The simplest profile is a uniform intensity. In this case, for an input amplitude of unity, the output amplitude for each polarization shall be $1/L_i$, where $L_i$ is the length of the structure in the $i^{th}$ dimension. A more appropriate output profile is one that is tapered, to improve beam quality and reduce side lobes. We selected a cosine-like field amplitude profile, similar to the electric field profile of the fundamental TE mode of the input waveguide.

Polarization control of the output field is of significant importance for properly interacting with the atomic vapor, especially the generation of high quality circular polarization. The polarization can be controlled using the phase between the two inputs. For linear polarization the phase shall be $n\pi$, with $n = 0, 1, 2, ...$, and for circular polarization the phase shall be $(n \pm 1/2)\pi$. Furthermore, for circular polarization, the amplitudes at each point are required to be the same for both orthogonal components. This is solved by using the design methodology described above, making the phase condition appropriate to guarantee the desired polarization. Another possible application of the polarization control is the ability to design linear polarization multiplexed field profiles, enabled by the relatively small cross coupling between the two orthogonal directions of propagations.

For optimal performance, additional design ingredients must be employed. One important ingredient is mitigation of reflections and scatterings from both edges of the coupling structure. Light entering the structure experiences an abrupt effective index change, causing scatterings in unwanted directions. These scatterings in turn cause interference patterns in the near field and affect the far field performance. In similar fashion, the light propagating in the structure encounters another abrupt transition at the far end of the structure. This transition causes both back reflections which interfere with the forward propagating field, and additional undesired scatterings. As a result, an increase in intensity can be seen in the near field, near the far edge of the structure, along with interference fringes. Fig. 4 illustrates these phenomena: The blue plot in Fig. 4a shows the simulated near field for the standard device, and Fig. 4b shows a side view of the simulated field. By introducing a gradual effective index change before and after the perturbed area (analog to impedance matching), these scatterings and reflections can be suppressed. This is realized by adding a few periods of ellipses with increasing radii, to gradually change the effective index from the pure waveguide effective index to the structure mean effective index. The simulated intensity for the matched case is seen in Fig. 4a as the red line, with side view in Fig. 4c. In this case four ellipses at each end were used to match the effective index. The ellipse diameters increase linearly, at both dimensions, from the smallest feature allowed in our lithography process (50 nm) to the required diameter at the first row. A longer device of 20 μm was used in this example to demonstrate this principle in more detail.

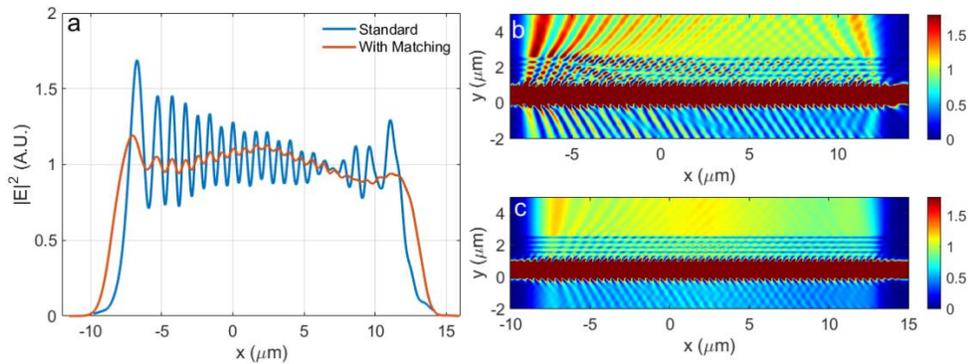

**Fig. 4.** Simulated field intensity for a 20 μm device: (a) At a distance of 2 μm in air above the structure, without matching (blue) and with four matching ellipses at each end (red); (b) side view of the non-matched structure; (c) side view of the matched structure.

## 3. Results

Due to computational limitations, all simulations were limited to 100x100 μm² sized structures. The simulation was constructed as follows. Two perpendicular 100 μm-wide waveguide mode sources, set to the fundamental TE mode of the waveguides, were used as inputs. A two-dimensional field monitor was placed several micrometers above the structure for recording the results. The $Si_3N_4$ slab thickness was 250 nm, with etch depth of 40 nm. As discussed above, the field intensity profile was predesigned to have a cosine-like profile for both propagating directions. The simulated near field result and calculated far field results for the 100x100 μm² device are shown in Fig. 5a, and c, respectively, with cross sections in Fig. 5b and d. The peak far field angle is at zero degrees, with a full-width half-maximum (FWHM) of 0.59 degrees.

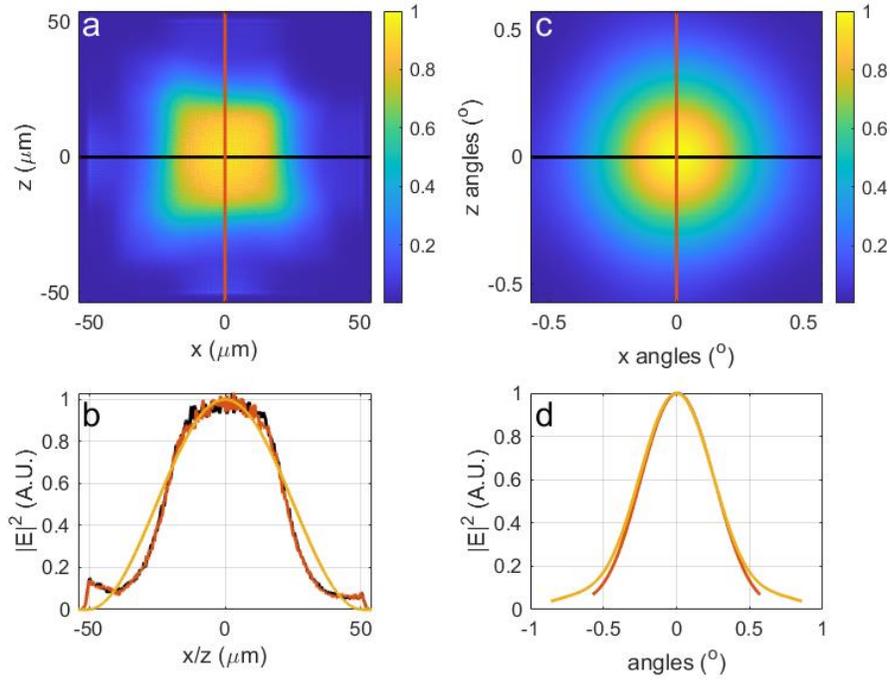

**Fig. 5.** (a) Simulated near field intensity of the 100x100 μm² quasi periodic structure; (b) Cross sections of near field intensities. Black: light propagating along z=0, red: light propagating along x=0, orange: target intensity; (c) Calculated far field intensity; (d) cross sections of the far field intensities. Colors are same as (b).

The circular polarization quality is now investigated. Left handed circular polarization (LCP) is created at the output field by assigning the x polarized input a $\pi/2$ phase, relative to the z polarized input. The polarization ellipse formed at each point in the field, by the two perpendicular fields, can be expressed using a basis of left and right handed circular polarizations. Using the formalism described by Santos et al. [25], we calculate the amplitudes of the two basis vectors, $a_+$ and $a_-$, at each point

$$a_+ = \frac{1}{2}\left(E_x^2 + 2E_xE_z \sin(\delta_x - \delta_z) + E_z^2\right)^{1/2} \qquad (6)$$

$$a_{-} = \frac{1}{2}\left(E_x^2 - 2E_x E_z \sin(\delta_x - \delta_z) + E_z^2\right)^{1/2} \tag{7}$$

Here $E_x$ and $E_z$ are the amplitudes of the x and z fields, and $\delta_x$ and $\delta_z$ are the initial phases of oscillation of the x and z fields. $a_{+}$ and $a_{-}$ are the amplitudes of the counterclockwise (left-handed) and clockwise (right-handed) rotating polarizations, respectively. The semimajor and semiminor axes of the polarization ellipse, $E_1$ and $E_2$, can be described by:

$$E_{1,2} = a_{+} \pm a_{-} \tag{8}$$

For pure LCP, $a_{-} = 0$ and $E_1 = E_2$ is expected. The circular polarization quality can be estimated by two metrics: the circular polarization contrast $a_{+}^2/a_{-}^2$ and the ratio between the two polarization ellipse axes $E_1/E_2$. Fig. 6 shows histograms of these two metrics for a beam area at its FWHM intensity. The mean circular polarization contrast is better than 30 dB and the mean $E_1/E_2$ ratio is 1.065.

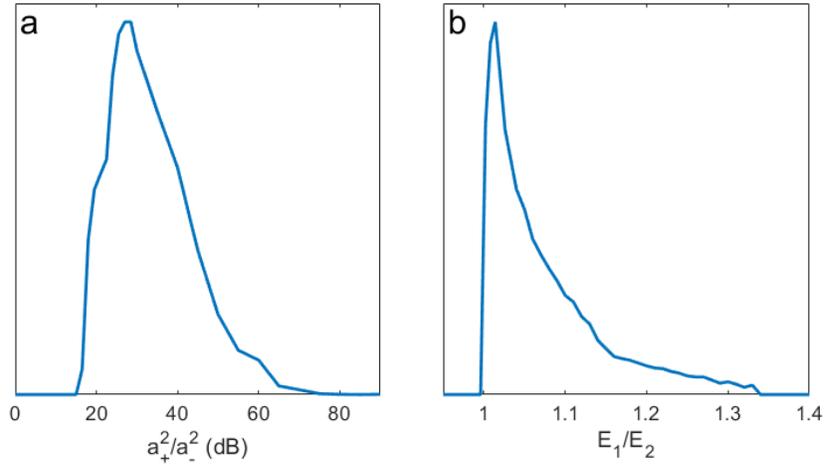

**Fig. 6.** Histogram of (a) the circular polarization contrast $a_{+}^2/a_{-}^2$ and (b) the ratio between the polarization ellipse axes $E_1/E_2$, for an area limited by the FWHM of the far field profile of the 100x100 µm² quasi periodic structure

We apply our design method to design different quasi periodic structures. This is for the purpose of pictorially illustrating the ability to design linear polarization multiplexed output field patterns, as well as visualizing the crosstalk between the fields emanated from the two orthogonally propagating modes. Crosstalk is defined as the intensity ratio between wanted and unwanted radiation in a specific location. Two 100x100 µm² structures were designed, both with the x-propagating, z-polarized light designed to show an image of the Technion logo, and the z-propagating, x-polarized light of the second structure was designed to an image of a star.

Fig. 7 shows the resulting simulated field intensities for these structures, exhibiting average crosstalk of 9.8 dB and 11 dB for the first and second structure, respectively.

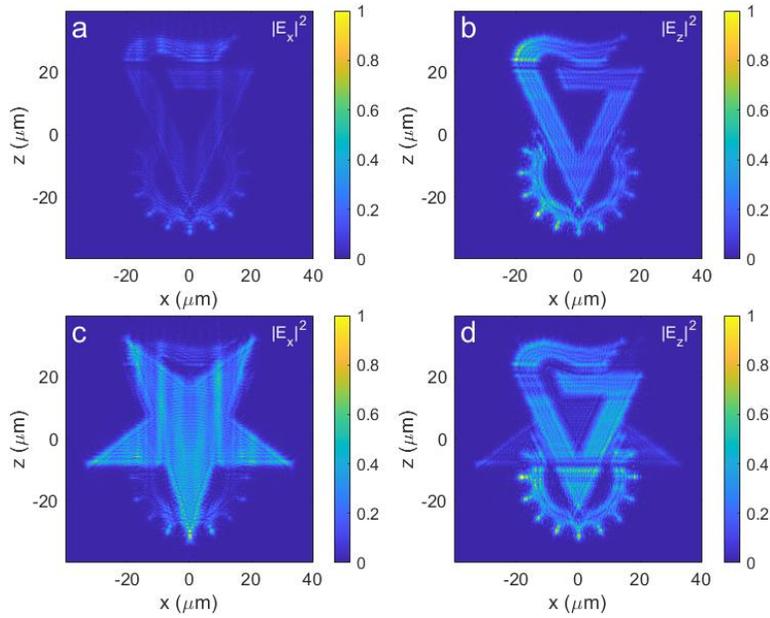

Fig. 7. Simulated field intensities with different designs for each polarization: (a) x and (b) z-polarized light for the structure designed to have only z-polarized light in the shape of the Technion logo; (c) z-propagating, x-polarized light, designed to show a star, and (d) x-propagating, z-polarized light, designed to show the Technion logo

## 4. Discussion

A new kind of a two dimensional quasi periodic structure was designed and simulated. These large structures enable control over the field amplitude, shape and polarization. The design method developed takes into account all aspects of the structure, some not usually accounted for in the simple one-dimensional model. In the case of the cosine-like profile, the simulated far field intensities showed good match to the design. The median circular polarization contrast at the far field, for a circle limited by the FWHM of the far field profile, is over 30 dB. Furthermore, we have shown the ability to multiplex the two linear polarizations with different field shapes, showing intensity separation of 11 dB.

Limitations of this method include two kinds of cross-talk between perpendicular polarizations. In the first kind, TE polarized light in the slab scatters into light polarized in the perpendicular direction in air. This occurs when light propagating through the structure is reflected in-plane by the curved ellipse outline. The perpendicularly propagating light is then out-coupled out of the slab and contributes to the perpendicular polarization cross-talk. The ratio of the main polarization to the perpendicular is over 30 dB in the structure shown here. The second kind of cross-talk is characterized by the inability to simultaneously match the required perturbation strength for both propagation directions at a specific point in the structure, which occurs for a more complicated beam shapes as was shown in Fig. 7. This kind of cross-talk cannot be completely eliminated, but can be mitigated by increasing the largest aspect ratio possible for the ellipses, thus decreasing the perturbation strength in the undesired polarization component. Unfortunately, due to fabrication constraints, this is not always possible.

The target total efficiencies of the structure presented here, for upwards and downwards radiation, are 7.9% and 6.6% respectively, with simulated results of 6.9% and 5.4%. The reason for setting a relatively low (less than 10%) efficiency target was the intention of using these 100x100 $\mu m^2$ arrays as a part of a larger beam generator to further approach the few hundred micrometer beam diameter, thus only a few percent of the input light should be emanated from each 100x100 μm$^2$ area. The structure dimensions shown in this paper are limited only by our computational capabilities. For the optimal design of 100x100 μm$^2$ structures, a deeper etch is required. The directionality of the quasi periodic structure can be improved by optimizing the buried oxide and cladding thickness [26,27], using two etch steps [20], or creating a double layer grating [28,29].

## 5. Conclusion

In summary, we have developed a method for the design of large quasi periodic structures for out-coupling light from the two-dimensional chip to free space. These structures enable control over field amplitude, shape, phase, and polarization for up to two simultaneous inputs. We have shown two examples for the use of these structures: circular polarization creation and linear polarization multiplexing. The circular polarization contrast obtained in simulation is over 30 dB for the 100x100 μm$^2$ structure, with dimensions limited only by computational power. We have shown linear polarization multiplexing by designing a structure with different field intensities for two linear polarizations, obtaining separation of 11 dB in simulation. Out results show this method can be used to successfully generate relatively large polarized optical beams directly from the integrated photonic circuit without 3D optical elements such as lenses, polarizers etc.

**Disclosures.** The authors declare no conflicts of interest.

**Data availability.** Data underlying the results presented in this paper are not publicly available at this time but may be obtained from the authors upon reasonable request.